\documentstyle[amsfonts,epsfig,12pt]{aipproc}
\begin{document}
\title{Solvable potentials, non-linear algebras, and
  associated coherent states} 
\author{F.\ Cannata\thanks{cannata@bo.infn.it}$^*$,
G.\ Junker\thanks{junker@theorie1.physik.uni-erlangen.de}$^{\dagger}$ and 
J.\ Trost\thanks{jtrost@theorie1.physik.uni-erlangen.de}$^{\dagger}$
}
%\\ 
\address{
$^*$Dipartmento di Fisica and INFN, Via Irnerio 46, I-40126 Bologna,
  Italy\\
$^{\dagger}$Institut f\"ur Theoretische Physik,
Universit\"at Erlangen-N\"urnberg,\\
Staudtstr.\ 7, D-91058 Erlangen, Germany}

%\lefthead{LEFT head}
%\righthead{RIGHT head}
\maketitle
%%%%%%%%%%%%%%%%%%%%%%%%%%%%%%%%%%%%%%%%%%%%%%%%%%%%%%%%%%%%%%%%%%%%%%%%%%%%

\begin{abstract}
Using the Darboux method and its relation with supersymmetric quantum
mechanics we construct all SUSY partners of the harmonic oscillator. With the
help of the SUSY transformation we introduce ladder operators for these
partner Hamiltonians and shown that they close a quadratic algebra. The
associated coherent states are constructed and discussed in some detail.
\end{abstract}
%%%%%%%%%%%%%%%%%%%%%%%%%%%%%%%%%%%%%%%%%%%%%%%%%%%%%%%%%%%%%%%%%%%%%%%%%%%%

\section*{Introduction}
Since the early days of quantum mechanics there has been enormous interest in
exactly solvable quantum systems. In fact, Schr\"odinger himself initiated a
program \cite{GJunker:Schr40} which resulted in the famous
Schr\"odinger-Infeld-Hull factorization method \cite{GJunker:InHu51}.
In the last 10-15 years this program has been revived in connection
with supersymmetric (SUSY) quantum mechanics \cite{GJunker:Jun1996}. To be a
little more precise, it 
has been found \cite{GJunker:Gen83} that the so-called property of
shape-invariance of a given 
Schr\"odinger potentials, which is in fact equivalent to the factorization
condition, is sufficient for the exact solvability of the eigenvalue problem
of the associated Schr\"odinger Hamiltonian. However, SUSY quantum mechanics
has also been shown to be an effective tool in finding new exactly solvable
systems. Here in essence one utilizes the fact that SUSY quantum mechanics
consists of a pair of essentially isospectral
Hamiltonians whose eigenstates are related by SUSY transformations. This is the
basic idea of a recent construction method for so-called conditionally exactly
solvable potentials \cite{GJunker:JuRo98}. Here one constructs a SUSY
quantum system for which, under 
certain conditions imposed on its parameters, one of the SUSY partner
Hamiltonians reduces to that of an exactly solvable (shape-invariant)
one. Other approaches, which are also based on the presence of pairs of
essentially isospectral Hamiltonians, go back to an idea formulated by Darboux
\cite{GJunker:Dar1882},
are based on the inverse scattering method \cite{GJunker:AbMo80}, 
or on the factorization method \cite{GJunker:Mil84}. Clearly, these
approaches are closely connected to each other and to the SUSY approach. 

In this paper we will construct with the help of the Darboux method all
possible SUSY partners of the harmonic oscillator Hamiltonian on the real line
and discuss their algebraic properties in some detail. In doing so we review
in the next section the Darboux method  and explicitly show its equivalence to
the supersymmetric approach. Section 3 then briefly presents the basic idea
for the construction of conditionally exactly solvable (CES)
potentials. Section 4 is devoted to a detailed discussion of the harmonic
oscillator case. Here we first present all possible SUSY partners of the
harmonic oscillator and give explicit expressions for the corresponding
eigenstates. Secondly, with the help of the standard ladder operators of the
harmonic oscillator we introduce similar ladder operators for the SUSY
partners and show that they close a quadratic algebra, which is also briefly
discussed. Finally, we introduce so-called non-linear coherent states which
are associated with this non-linear algebra. The properties of these coherent
states are discussed in some detail.
%%%%%%%%%%%%%%%%%%%%%%%%%%%%%%%%%%%%%%%%%%%%%%%%%%%%%%%%%%%%%%%%%%%%%%%%%%%%%

\section*{The Darboux method}
In this section we briefly review the Darboux method \cite{GJunker:Dar1882}
and show its connection to 
supersymmetric quantum mechanics \cite{GJunker:Jun1996}. In doing so we start
with considering a 
pair of standard Schr\"odinger Hamiltonians acting on $L^2({\mathbb R})$,
\begin{equation}
  \label{GJunker:2.1}
  H_{\pm}=-\frac{\hbar^2}{2m}\,\frac{\partial^2}{\partial x^2}+V_{\pm}(x)\;,
\end{equation}
and a linear operator
\begin{equation}
  \label{GJunker:2.2}
  A=\frac{\hbar}{\sqrt{2m}}\,\frac{\partial}{\partial x} + \Phi(x)\;,
   \quad \Phi:{\mathbb R}\to{\mathbb R}\;,
\end{equation}
obeying the intertwining relation
\begin{equation}
  \label{GJunker:2.3}
  H_+A=AH_-\;.
\end{equation}
It is obvious that this intertwining relation cannot be obeyed for arbitrary
functions $V_\pm$ and $\Phi$. In fact, the relation (\ref{GJunker:2.3})
explicitly reads 
\begin{equation}
  \label{GJunker:2.4}
  \begin{array}{r}
\displaystyle
\left(-\frac{\hbar^2}{2m}\,\Phi''(x)+V_+(x)\Phi(x)
 -\frac{\hbar}{\sqrt{2m}}\,V_-'(x)
 -\Phi(x)V_-(x)\right){\bf 1}=\qquad\qquad\qquad\\[4mm]
\displaystyle
\left(\frac{\hbar^2}{m}\,\Phi'(x)+\frac{\hbar}{\sqrt{2m}}\,V_-(x)
 -\frac{\hbar}{\sqrt{2m}}\,V_+(x) \right)\frac{\partial}{\partial x}\;.
  \end{array}
 \end{equation}
As the unit operator ${\bf 1}$ and the momentum operator (i.e.\
$\partial/\partial x$) are linearly independent, their coefficients have to
vanish. In other words,  
we are left with two conditions between the three functions $V_\pm$ and $\Phi$:
\begin{eqnarray}
 &&V_-(x)=V_+(x)-\frac{2\hbar}{\sqrt{2m}}\,\Phi'(x)\;, \label{GJunker:2.5}\\
 &&-\frac{\hbar^2}{2m}\,\Phi''(x)+V_+(x)\Phi(x)-
   \frac{\hbar}{\sqrt{2m}}\,V_-'(x)
   -\Phi(x)V_-(x)=0\;.\label{GJunker:2.6} 
\end{eqnarray}
Inserting the first one into the second one and integrating once we find
\begin{equation}
  \label{GJunker:2.7}
  \frac{\hbar}{\sqrt{2m}}\,\Phi'(x)-V_+(x)+\Phi^2(x)=-\varepsilon\;,
\end{equation}
where $\varepsilon$ is an arbitrary real integration constant
sometimes called factorization energy \cite{GJunker:Jun1996}. With this
relation and with 
(\ref{GJunker:2.5}) we can express the two potentials under consideration in
terms of the function $\Phi$:
\begin{equation}
  \label{GJunker:2.8}
  V_\pm(x)=\Phi^2(x)\pm\frac{\hbar}{\sqrt{2m}}\,\Phi'(x) +\varepsilon\;.
\end{equation}
At this point  one realizes that these are so-called SUSY partner
potentials \cite{GJunker:Jun1996}. In fact, using relations
(\ref{GJunker:2.8}) we note that 
\begin{equation}
  H_+=AA^\dagger+\varepsilon\;,\qquad H_-=A^\dagger A+\varepsilon\;.
\label{GJunker:2.9}
\end{equation}
These supersymmetric partner Hamiltonians are due to the intertwining relation
(\ref{GJunker:2.3}) essentially isospectral, that
is, 
\begin{equation}
\label{GJunker:2.10}
\mbox{spec\,}H_+\backslash \{\varepsilon\} = \mbox{spec\,}H_-\backslash
\{\varepsilon\} \;.
\end{equation}
Their eigenstates are related via SUSY transformations. To make
this more explicit, let us denote by $|\phi^\pm_n\rangle$ the eigenstates of
$H_\pm$ for eigenvalues $E_n>\varepsilon$,
\begin{equation}
  \label{GJunker:2.11}
  H_\pm|\phi^\pm_n\rangle=E_n|\phi^\pm_n\rangle\;.
\end{equation}
Then these states are related by SUSY transformations \cite{GJunker:Jun1996}
\begin{equation}
  \label{GJunker:2.12}
  |\phi^+_n\rangle=\frac{1}{\sqrt{E_n-\varepsilon}}\,A|\phi^-_n\rangle\;,\quad
  |\phi^-_n\rangle=\frac{1}{\sqrt{E_n-\varepsilon}}\,A^\dagger
                   |\phi^+_n\rangle\;. 
\end{equation}
In addition to the states in (\ref{GJunker:2.11}) one of the two Hamiltonians
$H_\pm$ may have an additional eigenstate $|\phi^\pm_\varepsilon\rangle$ with
eigenvalue $\varepsilon$ obeying the first-order differential equation
$A|\phi^-_\varepsilon\rangle=0$ and 
$A^\dagger|\phi^+_\varepsilon\rangle=0$,
respectively. In terms of the function $\Phi$ they explicitly
read
\begin{equation}
\label{GJunker:2.13}
\phi^\pm_\varepsilon(x)=N_\pm \exp\left\{\pm\frac{\sqrt{2m}}{\hbar}
\int dx\,\Phi(x)\right\}\;,
\end{equation}
where $N_\pm$ stands for a normalization constant. Clearly, only one of the
two solutions (\ref{GJunker:2.13}) may be square integrable. This situation
corresponds to an unbroken SUSY. If none of them is square integrable then
SUSY is said to be broken \cite{GJunker:Jun1996}.

The Darboux method reviewed in this section can now be used to find for a
given potential, say  $V_+$, all its possible SUSY partners $V_-$.
Firstly, one has to solve equation (\ref{GJunker:2.7}), that is,
finding all possible SUSY potentials $\Phi$. This in fact corresponds to find
all possible factorizations for the corresponding Hamiltonian $H_+$. Finally,
the corresponding SUSY partner $V_-$ can be obtained via (\ref{GJunker:2.5}).
In this way one can construct new exactly solvable potentials. The parameters
involved in the SUSY potential turn out to obey certain conditions and
therefore these new potentials are more precisely called conditionally exactly
solvable (CES) potentials. Let us note that the Darboux method may be
generalized to intertwining operators containing higher orders of the momentum
operator \cite{GJunker:AnIoSp93}.
%%%%%%%%%%%%%%%%%%%%%%%%%%%%%%%%%%%%%%%%%%%%%%%%%%%%%%%%%%%%%%%%%%%%%%%%%%%%

\section*{Modelling of CES potentials}
In this section we give some more details on the construction of CES
potentials using the Darboux method. As just mentioned above we start with a
given potential $V_+$ and try to find all its associated SUSY potentials. That
is, we have to find the most general solution of the generalized Riccati
equation (\ref{GJunker:2.7}). In doing so we will first linearize this
non-linear differential equation via the substitution
$\Phi(x)=(\hbar/\sqrt{2m})u'(x)/u(x)$,
\begin{equation}
  \label{GJunker:3.14}
  -\frac{\hbar^2}{2m}\,u''(x)+V_+(x)u(x)=\varepsilon u(x),
\end{equation}
which is actually a Schr\"odinger-like equation for $V_+$. Note, however, that
we are not restricted to normalizable solution of (\ref{GJunker:3.14}). In
other words, the energy-like parameter $\varepsilon$ is up to now still
arbitrary.  

In terms of $u$ the linear operator $A$ reads
\begin{equation}
\label{GJunker:3.15}
A=\frac{\hbar}{\sqrt{2m}}
\left(\frac{\partial}{\partial x} + \frac{u'(x)}{u(x)}\right)
\end{equation}
and thus is only a well-defined operator on $L^2({\mathbb R})$ if $u$ does not
have any zeros on the real line. As a consequence we may admit only those
solutions of (\ref{GJunker:3.14}) which have no zeros. Form Sturmian theory we
know that 
this is only possible if $\varepsilon$ is below the ground-state energy of
$H_+$ which we will denote by $E_0$. Hence, we obtain a first condition on the
parameter $\varepsilon$, which reads $\varepsilon < E_0$. This also implies
that $\varepsilon$ does not belong to the spectrum of $H_+$. In fact, the
associated eigenfunction (\ref{GJunker:2.13}) would read $\phi^+_\varepsilon
(x)= N_+ u(x)$, which is not normalizable due to condition put on
$\varepsilon$.  

The above condition on $\varepsilon$ is still not sufficient to guarantee a
nodeless solution. Being a second-order linear differential equation
(\ref{GJunker:3.14}) has two linearly independent fundamental solutions denoted
by $u_1$ and $u_2$. Hence, the most general solution for $\varepsilon < E_0$
is given by a linear combination of the fundamental ones:
\begin{equation}
\label{GJunker:3.16}
u(x)=\alpha \, u_1(x) +\beta \, u_2(x)\;.
\end{equation}
Therefore, the condition that $u$ does not vanish also imposes conditions on
the parameters $\alpha$ and $\beta$, which have to be studied case by case
\cite{GJunker:JuRo98}. 

Let us now assume that $H_+$ is an exactly solvable Hamiltonian, which means
that its eigenvalues $E_n$ and eigenstates $|\phi^+_n\rangle$ are exactly known
in closed form. For simplicity we have assumed that $H_+$ has a purely discrete
spectrum enumerated by $n=0,1,2,\ldots$ such that
$\varepsilon<E_0<E_1<\ldots$. Then via the method outlined above one can
construct all its SUSY partners $H_-$ which are conditionally exactly solvable
due to the conditions which have to be imposed on the parameters
$\alpha,\beta$ and $\varepsilon$. By construction the eigenvalues of $H_+$ are
also eigenvalues of $H_-$ and the corresponding eigenfunctions are obtained via
the SUSY transformation (\ref{GJunker:2.12}). In the case of unbroken SUSY
$H_-$ has one additional eigenvalue $\varepsilon$ which belongs to its ground
state given by $\phi^-_\varepsilon(x)=N_-/u(x)$. Finally, we note that in
terms of $u$ the partner potentials read
\begin{equation}
  \label{GJunker:3.17}
V_-(x)=\frac{\hbar^2}{m}\left(\frac{u'(x)}{u(x)}\right)^2
               -V_+(x)+2\varepsilon
\end{equation}
and form a two-parameter family label by $\varepsilon$
and $\beta/\alpha$. Note that only the quotient $\beta/\alpha$ or its inverse
is relevant for (\ref{GJunker:3.17}). For various examples of CES potentials
found by this method see \cite{GJunker:JuRo98}. Here we limit our discussion
to those related to the harmonic oscillator.
%%%%%%%%%%%%%%%%%%%%%%%%%%%%%%%%%%%%%%%%%%%%%%%%%%%%%%%%%%%%%%%%%%%%%%%%%%%

\section*{The harmonic oscillator}
In this section we will now construct all possible SUSY partner potentials for
the harmonic oscillator $V_+(x)=(m/2)\omega^2 x^2$, $\omega >0$, via the
Darboux method. The corresponding 
Schr\"odinger-like equation (\ref{GJunker:3.14}) reads in this 
case\footnote{From now on we will use dimensionless quantities, that is, $x$
  is given in units of $\sqrt{\hbar/m\omega}$ and all energy-like quantities
  are given in units of $\hbar \omega$.}
\begin{equation}
  \label{GJunker:4.18}
  -\frac{1}{2}\,u''(x)+\frac{1}{2}\,x^2u(x)=\varepsilon u(x)
\end{equation}
and has as general solution a linear combination of confluent hypergeometric
functions 
\begin{equation}
  \label{GJunker:4.19} \textstyle
   u(x)= e^{-x^2/2}\left[
    \alpha\;_{1}F_{1}(\frac{1-2\varepsilon}{4},\frac{1}{2},x^2) +
    \beta\; x\;_{1}F_{1}(\frac{3-2\varepsilon}{4},\frac{3}{2},x^2)\right]\;.
\end{equation}
The condition that $u$ does not have a real zero implies that $\alpha$ must
not vanish and thus can be set equal to unity without loss of
generality. Furthermore, $\beta$ has to obey the inequality
\cite{GJunker:JuRo98,GJunker:CaJuTr98} 
\begin{equation}
  \label{GJunker:4.20}
|\beta|<\beta_c(\varepsilon):= 2\,
\frac{\Gamma(\frac{3}{4}-\frac{\varepsilon}{2})}
     {\Gamma(\frac{1}{4}-\frac{\varepsilon}{2})}\;.
\end{equation}
The corresponding partner potentials of the harmonic oscillator then read
according to (\ref{GJunker:3.17})
\begin{equation}
  \label{GJunker:4.21}
  V_-(x)=\left(\frac{u'(x)}{u(x)}\right)^2
               -\frac{1}{2}\,x^2+2\varepsilon\;.
\end{equation}
We note that for the above $u$ SUSY remains unbroken and therefore,
the spectral properties of $H_-$ are given by 
\begin{equation}
  \begin{array}{l}
   {\mathrm spec}\,H_- = \{\varepsilon,E_0,E_1,\ldots\}\quad\mbox{with}\quad
  E_n=n+\frac{1}{2}\;,\quad n=0,1,2,\ldots\;,\\[2mm]
\displaystyle
  \phi^-_\varepsilon(x) =\frac{N_-\, e^{x^2/2}}
    {_{1}F_{1}(\frac{1-2\varepsilon}{4},\frac{1}{2},x^2) +
    \beta\; x\;_{1}F_{1}(\frac{3-2\varepsilon}{4},\frac{3}{2},x^2)}\;,\\[4mm]
\displaystyle
\phi^-_n(x)=\frac{\exp\{-x^2/2\}}{[\sqrt{\pi}\,2^{n+1} n!
 (n+1/2-\varepsilon)]^{1/2}}
 \left[H_{n+1}(x)+\left(\frac{u'(x)}{u(x)}-x\right)H_n(x)\right]\;,
  \end{array}
\end{equation}
where $H_n$ denotes the Hermite polynomial of degree $n$.
Figures of the potential family (\ref{GJunker:4.21}) for various values of
$\varepsilon$ and $\beta$ can be found in \cite{GJunker:JuRo98}. Here let us
stress that one can even allow for complex valued $\beta\in{\mathbb
  C}\backslash [-\beta_c(\varepsilon),\beta_c(\varepsilon)]$ which in turn
will give rise to complex potentials generating the same real spectrum
\cite{GJunker:CaJuTr98}. We also note that the present CES potential
(\ref{GJunker:4.21}) contains as special cases those
previously obtain by Abraham and Moses \cite{GJunker:AbMo80} and by Mielnik
\cite{GJunker:Mil84}. See also \cite{GJunker:JuRo98} for a detailed discussion.
%%%%%%%%%%%%%%%%%%%%%%%%%%%%%%%%%%%%%%%%%%%%%%%%%%%%%%%%%%%%%%%%%%%%%%%%%%%%

\subsection*{Algebraic Structure}
We will now analyse the algebraic structure for the partner Hamiltonians of
the harmonic oscillator. In fact, using the standard raising and lowering
operators of the harmonic oscillator 
$H_+=AA^\dagger+\varepsilon=a^\dagger a + 1/2$, 
\begin{equation}
  \label{GJunker:4.22}
  a=\frac{1}{\sqrt{2}}\left(\frac{\partial}{\partial x} + x\right)\;,\quad
  a^\dagger=\frac{1}{\sqrt{2}}\left(-\frac{\partial}{\partial x} + x\right)\;,
\end{equation}
which close the linear algebra
\begin{equation}
  \label{GJunker:4.23}
[H_+,a]=-a\;,\quad [H_+,a^\dagger ]=a^\dagger\;,\quad [a,a^\dagger]={\bf 1}\;,
\end{equation}
one may introduce via the SUSY transformation (\ref{GJunker:2.12}) similar
ladder operators for the SUSY partners \cite{GJunker:JuRo97}
\begin{equation}
  \label{GJunker:4.24}
  B=A^\dagger a A\;,\quad B^\dagger =A^\dagger a^\dagger A\;,
\end{equation}
which act on the eigenstates of $H_-$ in the following way
\begin{equation}
\label{GJunker:4.25}
  \begin{array}{l}
  B|\phi^-_{n+1}\rangle=
   \sqrt{(n+\frac{1}{2}-\varepsilon)(n+1)(n+\frac{3}{2}-\varepsilon)}
    |\phi^-_{n}\rangle\;,\\[2mm]
 B^\dagger|\phi^-_{n}\rangle=
   \sqrt{(n+\frac{3}{2}-\varepsilon)(n+1)(n+\frac{1}{2}-\varepsilon)}
|\phi^-_{n+1}\rangle\;,\\[2mm]
B|\phi^-_{0}\rangle=0\;,\quad  B|\phi^-_{\varepsilon}\rangle=0\;,\quad 
B^\dagger|\phi^-_{\varepsilon}\rangle=0\;.
  \end{array}
\end{equation}
The last two relations explicate that the ground state
$|\phi^-_{\varepsilon}\rangle$ of $H_-$ is isolated in
the sense that it cannot be reached via $B$ from any of the excited states and,
vice versa, the excited states cannot be constructed with $B^\dagger$ from
$|\phi^-_{\varepsilon}\rangle$. 
These ladder operators close together with the Hamiltonian $H_-$ the
quadratic, hence non-linear, algebra
\begin{equation}
  \label{GJunker:4.27}
[H_-,B]=-B\;,\quad [H_-,B^\dagger]=B^\dagger\;,\quad 
[B,B^\dagger]=3H_-^2-4\varepsilon H_-+\varepsilon^2\;.
\end{equation}
This quadratic algebra belongs to the class of so-called $W_2$ algebras and
may be viewed as a polynomial deformation of the $su(1,1)$ Lie algebra. Such
deformations have been discussed by Ro\u{c}ek \cite{GJunker:Ro91} and, within
a more general context, by Karassiov \cite{GJunker:Ka94} and Katriel and
Quesne \cite{GJunker:KaQu96}. The quadratic Casimir operator associated with
the algebra (\ref{GJunker:4.27}) reads
\begin{equation}
C = BB^{\dag} - \Psi(H_{-})\;,\quad
\Psi(H_{-}) - \Psi(H_{-}-1) = 3H^{2}_{-} - 4\varepsilon H_{-} + 
\varepsilon^{2} \;.
\label{GJunker:4.29}
\end{equation}
In the Fock space representation (\ref{GJunker:4.25}) we have the following
explicit expression
\begin{equation}
\Psi(H_{-}) = (H_{-}-\varepsilon)(H_{-}+{\textstyle\frac{1}{2}})
(H_{-}+1-\varepsilon)
\label{GJunker:4.30}
\end{equation}
and the relations $BB^{\dagger}=\Psi(H_{-})$ and $B^{\dagger}B=
\Psi(H_{-}-1)$. Hence the Casimir (\ref{GJunker:4.29}) vanishes within this
representation as 
expected \cite{GJunker:Ka94,GJunker:KaQu96}. 
%%%%%%%%%%%%%%%%%%%%%%%%%%%%%%%%%%%%%%%%%%%%%%%%%%%%%%%%%%%%%%%%%%%%%%%%%%%%

\subsection*{Non-linear coherent states}
Let us now construct the non-linear coherent states \cite{GJunker:JuRo98a}
associated with the quadratic algebra (\ref{GJunker:4.27}). There are several
ways to define such states \cite{GJunker:ZhFeGi90}. Here we will define them
as eigenstates 
of the ``non-linear'' annihilation operator $B$, leading essentially to
so-called Barut-Girardello coherent states \cite{GJunker:BaGi71}. We also note
that the construction procedure presented below is very similar to that of
coherent states associated with quantum groups \cite{GJunker:Spi95}.

Let us note that the ground state $|\phi^-_\varepsilon\rangle$ of $H_-$ is
isolated and therefore we may construct the coherent states over the excited
states $\{|\phi^-_n\rangle\}_{n\in{\mathbb N}_0}$ only. For this reason we make
the ansatz 
\begin{equation}
|\mu\rangle = \sum_{n=0}^{\infty} c_{n} \, \mu^{n} \,  |\phi^-_n\rangle \;,
\label{GJunker:4.31}
\end{equation}
where $\mu$ is an arbitrary complex number and the real coefficients $c_n$ are
to be determined from the defining relation
\begin{equation}
\label{GJunker:4.32}
B|\mu\rangle = \mu \, |\mu\rangle = \sum_{n=0}^\infty c_n\,\mu^{n}\,
B|\phi^-_n\rangle \;.
\end{equation}
Using relations (\ref{GJunker:4.25}) we obtain the following
recurrence relation for the $c_{n}$'s,
\begin{equation}
c_{n+1} = c_{n}\left[\textstyle(n+\frac{1}{2}-\varepsilon) (n+1)
(n+\frac{3}{2}-\varepsilon)\right]^{-1/2}  \;.
\label{GJunker:4.33}
\end{equation}
That is, the coefficients $c_{n}$ for $n\geq1$ can be expressed in terms of
$c_{0}$, 
\begin{equation}
c_{n} = c_{0}\left[\textstyle n! (\frac{1}{2}-\varepsilon)_{n}
(\frac{3}{2}-\varepsilon)_{n}\right]^{-1/2} 
\label{GJunker:4.34}
\end{equation}
where $(z)_{n}=\Gamma(z+n)/\Gamma(z)$ denotes Pochhammer's symbol. The
remaining coefficient $c_{0}=c_{0}(\mu)$ is determined via the normalization
of the coherent states
\begin{equation}
\label{GJunker:4.35}
\langle\mu|\mu\rangle = 
c_{0}^{2}(\mu) \,  \sum_{n=0}^{\infty}
\frac{|\mu|^{2n}}{n!} \frac{1}{(\frac{1}{2}-\varepsilon)_{n}
(\frac{3}{2}-\varepsilon)_{n}} = 1 \; . 
\end{equation}  
Thus, we can express $c_{0}$ in terms of a generalized
hypergeometric function \cite{GJunker:Erd53}
\begin{equation}
c_{0}^{-2}(\mu) =
\ _{0}F_{2}\left({\textstyle\frac{1}{2}}-\varepsilon,
{\textstyle\frac{3}{2}} - \varepsilon;
|\mu|^{2}\right) \; . 
\label{GJunker:4.36}
\end{equation}

Let us now discuss some properties of these non-linear coherent states. First
we note that these states are not orthogonal for $\mu\neq\nu$ as expected:
\begin{equation}
\label{GJunker:4.37}
\langle\mu|\nu\rangle = c_{0}(\mu) \,  c_{0}(\nu) \ 
_{0}F_{2}\left({\textstyle\frac{1}{2}}-\varepsilon,
{\textstyle\frac{3}{2}} -\varepsilon;
\mu^{*}\nu\right) \; . 
\end{equation}
Secondly, let us investigate whether these states form an overcomplete set. In
other words, we consider the question: Can these states generate a resolution
of the unit operator? For this we have to recall that the non-linear
coherent states have been constructed over the excited states of $H_-$.
Therefore, we start with postulating a positive measure $\rho$ on
the complex $\mu$-plane obeying the following resolution of unity:
\begin{equation}
\int_{\mathbb C} d\rho(\mu^{*},\mu)\, |\mu\rangle\langle\mu| = {\bf 1} -
|\phi^-_\varepsilon\rangle\langle \phi^-_\varepsilon| \;.
\label{GJunker:4.38}
\end{equation}
Within the polar decomposition $\mu=\sqrt{x}\,e^{i\varphi}$ we make
the ansatz 
\begin{equation}
d\rho(\mu^{*},\mu) = \frac{d\varphi \, dx \,
\sigma(x)}{2\pi c_{0}^{2}(\sqrt{x})} \ ,
\label{GJunker:4.39}
\end{equation}
with a yet unknown positive density $\sigma$ on the positive
half-line. Inserting this ansatz into (\ref{GJunker:4.38}) we obtain the
following conditions on $\sigma$
\begin{equation}
\int_{0}^{\infty} dx\, \sigma(x) \, x^{n} = \Gamma(n+1)
 \, \frac{\Gamma(\frac{1}{2}-\varepsilon+n)
\Gamma(\frac{3}{2}-\varepsilon+n)}
 {\Gamma(\frac{1}{2}-\varepsilon)\Gamma(\frac{3}{2}-\varepsilon)}\;, \quad
 n=0,1,2,\ldots \;.  
\label{GJunker:4.40}
\end{equation}
Hence, $\sigma$ is a probability density on the positive half-line defined by
its moments given on the right-hand side of (\ref{GJunker:4.40}). 
Let us note that the integral in (\ref{GJunker:4.40}) may be viewed as a
Mellin transformation \cite{GJunker:Erd54} of $\sigma$ and in turn the latter
is given by the inverse Mellin transformation of the moments. This
inverse Mellin transformation turns out to lead to the integral representation
of Meijer's G-function \cite{GJunker:Erd53}. In other words, we have the
explicit form:
\begin{equation}
\sigma(x) =
 \frac{1}{\Gamma(\frac{1}{2}-\varepsilon) \Gamma(\frac{3}{2}-\varepsilon)} 
\, G^{30}_{03}
 \left(x|0,-{\textstyle\frac{1}{2}}-\varepsilon, 
 {\textstyle\frac{1}{2}}-\varepsilon \right) \;.
\end{equation}
\begin{figure}[t]%fig 1
\vspace{100mm}%
\includegraphics{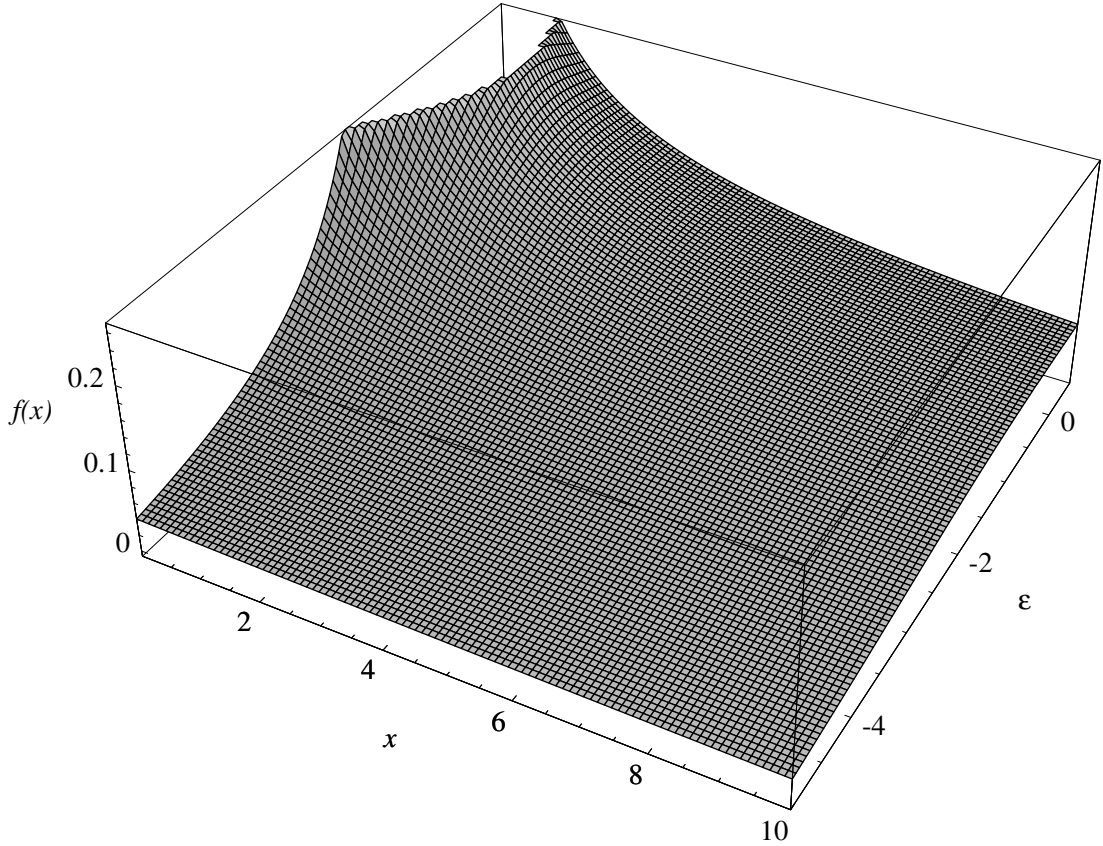}%
\vspace{20mm}%
\caption{The radial density $f(x)=\sigma(x)/c_0^2(\sqrt{x})$ giving rise to
 the resolution of unity (\ref{GJunker:4.38}) with (\ref{GJunker:4.39}) as a
 function of $x=|\mu|^2$ and for various parameters
 $\varepsilon<\frac{1}{2}$.}% 
\end{figure}%
In Figure 1 a plot of the radial density $f(|\mu|^2)=2\pi\,d\rho(\mu^*,\mu)/
(d\varphi d|\mu|^2)$ is given showing that it leads to a well-behaved
positive measure on the complex $\mu$-plane.

Finally, let us point out that similar non-linear coherent states associated
with the CES potentials of the radial harmonic oscillator have been
constructed in \cite{GJunker:JuRo98a}. In that case broken as well as unbroken
SUSY can be considered and the corresponding symmetry algebra is a cubic
one. In analogy to the discussion in \cite{GJunker:JuRo98a} one can show that
the coherent states discussed here are also minimum uncertainty states.
%%%%%%%%%%%%%%%%%%%%%%%%%%%%%%%%%%%%%%%%%%%%%%%%%%%%%%%%%%%%%%%%%%%%%%%%%%%%%%

\section*{Acknowledgements}
One of us (G.J.) would like to thank the organizers for their kind
invitation to this very stimulating meeting. In particular,
he has enjoyed valuable discussions with C.M.\ Bender, P.P.\ Kulish and A.\
Odzijewicz during this conference. 
%%%%%%%%%%%%%%%%%%%%%%%%%%%%%%%%%%%%%%%%%%%%%%%%%%%%%%%%%%%%%%%%%%%%%%%%%%%%

%%%%%%%%%%%%%%%%%%%%%%%%%%%%%%%%%%%%%%%%%%%%%%%%%%%%%%%%%%%%%%%%%%%%%%%%%%%%


\begin{references}
\bibitem{GJunker:Schr40} Schr\"odinger E., {\it Proc.\ Roy.\ Irish Acad.} {\bf
    46A}, 9--14 (1940); {\bf 46A}, 183--206 (1941); {\bf 47A}, 53--54 (1941).
\bibitem{GJunker:InHu51} Infeld L.\ and Hull T.E., {\it Rev.\ Mod.\ Phys.}
    {\bf 23}, 21--68 (1951).
\bibitem{GJunker:Jun1996}Junker G., {\it Supersymmetric Methods in Quantum and
    Statistical Physics}, Berlin: Springer-Verlag, 1996.
\bibitem{GJunker:Gen83} Gendenshte\^{i}n L.\'{E}., {\it JETP Lett.} {\bf 38},
    356--358 (1983).
\bibitem{GJunker:JuRo98} Junker G.\ and Roy P., {\it Conditionally exactly
    solvable potentials: A supersymmetric construction method}, preprint
    quant-ph/9803024. 
\bibitem{GJunker:Dar1882}Darboux G., {\it Comptes Rendus Acad.\ Sci.\ (Paris)}
    {\bf 94}, 1456--1459 (1882).
\bibitem{GJunker:AbMo80}Abraham P.B.\ and Moses H.E., {\it Phys.\ Rev.\ A}
    {\bf 22}, 1333--1686 (1980).\\
    Luban M.\ and Pursey D.L., {\it Phys.\ Rev.\ D} {\bf 33}, 431--436
    (1986).\\
    Pursey D.L., {\it Phys.\ Rev.\ D} {\bf 33}, 1048--1055 (1986).
\bibitem{GJunker:Mil84}Mielnik B., {\it J.\ Math.\ Phys.} {\bf 25}, 3387--3389
    (1984). 
\bibitem{GJunker:AnIoSp93}Andrianov A.A., Ioffe M.V.\ and Spiridonov V.P.,
    {\it Phys.\ Lett.\ A} {\bf 174}, 273--179 (1993).\\
    Andrianov A.A., Ioffe M.V., Cannata F.\ and Dedonder J.-P., {\it Int.\ J.\
    Mod.\ Phys.\ A} {\bf 10}, 2683--2702 (1995).\\
    Samsonov B.F., {\it J.\ Math.\ Phys.\ A} {\bf 28}, 6989--6998 (1995).
\bibitem{GJunker:CaJuTr98} Cannata F., Junker G.\ and Trost J., {\it
    Schr\"odinger operators with complex potentials but real spectrum},
    preprint quant-ph/9805085. 
\bibitem{GJunker:JuRo97}Junker G.\ and Roy P., {\it Phys. Lett.\ A} {\bf 232},
    155--161 (1997).\\
    Junker G.\ and Roy P., {\it Supersymmetric construction
    of exactly solvable potentials and non-linear algebras}, preprint
    quant-ph/9709021. 
\bibitem{GJunker:Ro91} Ro\u{c}ek M., {\it Phys. Lett.\ B} {\bf 255}, 554--557
    (1991).
\bibitem{GJunker:Ka94} Karassiov V.P., {\it J.\ Phys.\ A} {\bf 27}, 153--165
    (1994). 
\bibitem{GJunker:KaQu96} Katriel J.\ and Quesne C., {\it J.\ Math.\ Phys.}
    {\bf 37}, 1650--1661 (1996).
\bibitem{GJunker:JuRo98a} Junker G.\ and Roy P., {\it Non-linear coherent
    states associated with conditionally exactly solvable problems}, preprint
    (1998). 
\bibitem{GJunker:ZhFeGi90}Zhang W.-M., Feng D.H.\ and Gilmore R., {\it Rev.\
    Mod.\ Phys.} {\bf 62}, 867--927 (1990).
\bibitem{GJunker:BaGi71} Barut A.O.\ and Girardello L., {\it Commun.\ Math.\
    Phys.} {\bf 21}, 41--55 (1971).
\bibitem{GJunker:Spi95} Spiridonov V., {\it Phys.\ Rev.\ A} {\bf 52},
    1909--1935 (1995).\\
    Odzijewicz A., {\it Commun.\ Math.\ Phys.} {\bf 192}, 183--215 (1998).
\bibitem{GJunker:Erd53} Erd\'elyi A., Magnus W., Oberhettinger F.\ and Tricomi
    F.G.,  {\it Higher Transcedental Functions}, Volume I, New York:
    McGraw-Hill, 1953. 
\bibitem{GJunker:Erd54} Erd\'elyi A., Magnus W., Oberhettinger F.\ and Tricomi
    F.G.,  {\it Tables of Integral Transforms}, Volume I, New York:
    McGraw-Hill, 1954.
\end{references}
\end{document}